\documentclass[preprint,12pt]{elsarticle}

\usepackage{amsmath,amsfonts}
\usepackage{algorithm}
\usepackage{algpseudocode}
\usepackage{xurl}
\usepackage{hyperref}
\usepackage{array}
\usepackage[caption=false,font=normalsize,labelfont=sf,textfont=sf]{subfig}
\usepackage{textcomp}
\usepackage{stfloats}
\usepackage{url}
\usepackage{verbatim}
\usepackage{graphicx}
\usepackage{orcidlink}

\usepackage{soul}
\sethlcolor{yellow}
\soulregister\cite7
\soulregister\ref7
\soulregister\eqref7
\soulregister\pageref7
\soulregister\texttt7
\soulregister\emph7

\usepackage{array}
\usepackage{tabularx}
\usepackage{ragged2e}
\usepackage{arydshln}
\journal{EAAI}

\begin{document}

\begin{frontmatter}



\title{Minimum Bisection Problem: Machine Learning–Based Penalty Parameter Tuning for Optimization on Quantum Annealers}


\author{Renáta Rusnáková\corref{cor1}\,\orcidlink{0009-0004-3966-2201}}
\author{Martin Chovanec\,\orcidlink{0000-0001-9640-6491}}
\author{Juraj Gazda\,\orcidlink{0000-0002-7334-9540}}
\cortext[cor1]{Corresponding author: R. Rusnáková (e-mail: renata.rusnakova@tuke.sk).}
\affiliation{organization={Department of Computers and Informatics, Faculty of Electrical Engineering and Informatics, Technical University of Košice},
           city={Košice},
           postcode={04200},
           country={Slovakia}}

%% Abstract
\begin{abstract}
The Minimum Bisection Problem is a fundamental computationally hard graph
partitioning problem with applications in parallel computing, network design, and large-scale data processing. When formulated as a
Quadratic Unconstrained Binary Optimization problem for quantum
annealing, solution quality depends critically on the penalty parameter
that enforces balanced partitions. Selecting this parameter is
problem-dependent and typically relies on manual tuning or heuristics.
This paper proposes a machine learning-based approach for automatic
penalty-parameter tuning developed specifically for the Minimum
Bisection Problem. We first derive a graph-dependent initial penalty
estimate and then use two Gradient Boosting Regressor models to predict the
endpoints of an effective penalty-multiplier interval from the number of
nodes, graph density, and the initial estimate. The final penalty is
obtained from the predicted interval and used to construct the model
solved by D-Wave's quantum annealing solvers. The models were calibrated on 607 Erd\H{o}s--R\'enyi graphs, with Metis and Kernighan--Lin as classical references, and evaluated on 126 independently generated instances with up to 4000 nodes. Under the adopted
experimental setup, the predicted penalties enabled the hybrid solver
to return balanced partitions for all evaluation instances and lower
cut values than Metis in every case.
\end{abstract}

\begin{graphicalabstract}
\begin{figure}
    \centering
    \includegraphics[width=1\linewidth]{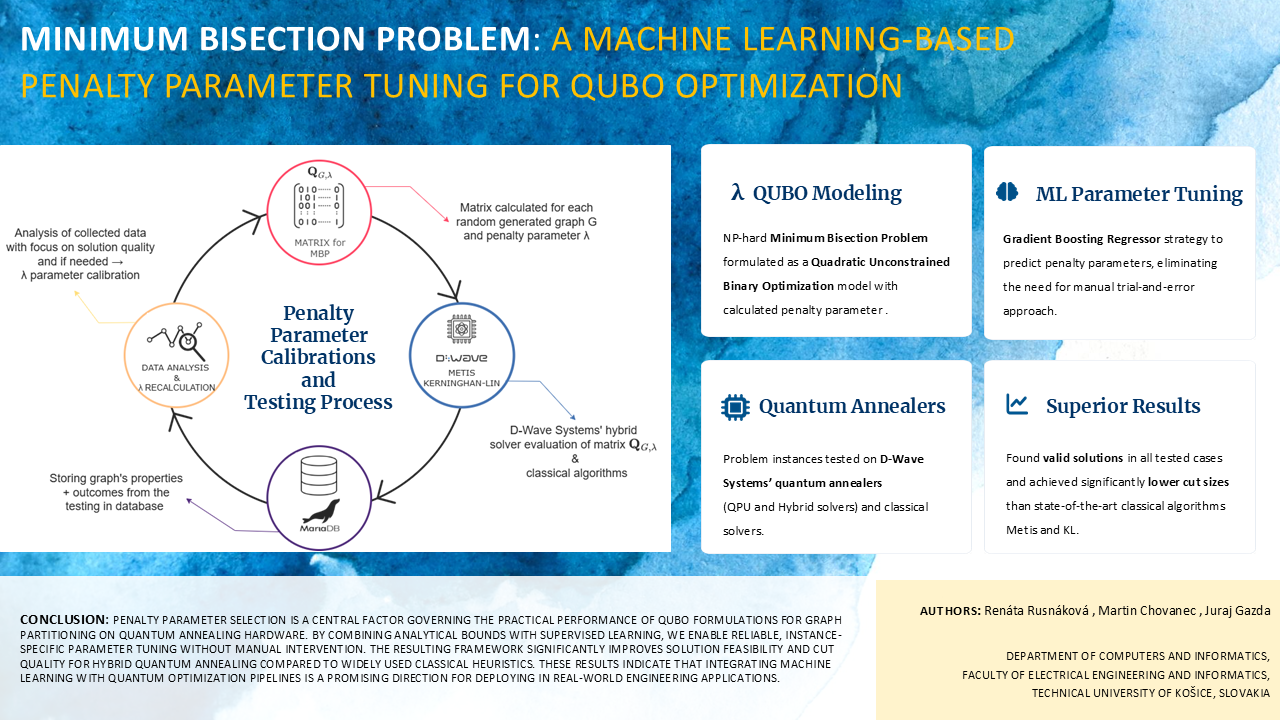}
    \label{fig:placeholder}
\end{figure}
\end{graphicalabstract}

\begin{highlights}
\item Machine learning automates QUBO penalty tuning for balanced graph partitioning
\item Two regression models predict a suitable penalty-parameter interval
\item Data-driven penalty selection improves feasibility and partition quality
\end{highlights}

%% Keywords

\end{frontmatter}



\section{Introduction}
\label{sec:Introduction}

Graph partitioning is a fundamental problem in combinatorial optimization with applications in multiprocessor scheduling, microchip design, large-scale network analysis, machine learning, and transportation systems. In these areas, partitioning is used to divide a complex system into smaller, weakly connected parts, which can reduce communication costs, distribute computational workloads, and improve scalability.

Given an undirected graph, the graph partitioning problem divides the node set into two or more subsets while minimizing the number or total weight of edges connecting different subsets. An important constrained variant is the Minimum Bisection Problem (MBP), in which the graph must be divided into two subsets of equal size while minimizing the cut between them~\cite{Karpinski2002}.

The MBP is NP-hard~\cite{Kucera1955,Garey1979}, and exact methods become computationally demanding as the size of the graph increases~\cite{Quinton2024}. For this reason, large instances are usually solved using approximation algorithms and heuristics~\cite{Buluc2015,Sanders2011,Barnard1993,Hendrickson1995,Sanders2020,Akhremtsev2018}. Common approaches include multilevel partitioning methods such as Metis~\cite{Karypis1998,Karypis1997} and local-refinement algorithms such as Kernighan--Lin~\cite{Kernighan1970,Patil2021}. Their performance depends on the structure of the graph, the initialization, and how strictly the balance constraint is enforced.

The MBP can be written as a Quadratic Unconstrained Binary Optimization (QUBO) problem~\cite{Lucas2014,Glover2019}. In this formulation, the number of edges crossing between the two subsets is represented by the objective term, while a quadratic penalty term is added to enforce equal partition sizes. The resulting QUBO can be solved by classical optimization methods, quantum-inspired algorithms, or quantum annealers.

D-Wave Systems provides access to quantum processing unit (QPU) hardware, as well
as hybrid quantum--classical solvers for binary quadratic models~\cite{Dwave_Hybrid}. The hybrid solvers combine classical processing with quantum sampling and can handle larger problems than those that can currently be embedded directly on a QPU~\cite{Raymond2023}. In this study, the hybrid solver is used for large graph instances of MBP, while direct-QPU experiments are performed only for smaller graphs.

A central challenge in QUBO modeling is the selection of penalty parameters. If a penalty is too small, the solver may return solutions that violate the imposed constraints; if it is excessively large, the penalty term may dominate the objective coefficients and reduce the solver's ability to distinguish solutions according to their original objective values. A range of approaches has therefore been proposed to determine penalty weights more systematically. 
Earlier studies investigated static, exact, sequential, and automatically
generated penalty weights for different classes of QUBO problems
~\cite{Ayodele2022,GarciaAyodeleMoraglio2022,LiuMoraglio2025}. Recent work has extended this research in several directions. Alessandroni \textit{et al.} examined the quantum Big-$M$ problem and proposed tighter penalty weights to prevent the penalty terms from dominating the original objective function~\cite{Alessandroni2025}. Liu and Moraglio introduced a structure-aware
method for scaling multiple penalty weights according to the structure of
the QUBO formulation~\cite{LiuMoraglio2026}, while Alessandroni
\textit{et al.} later proposed a scalable procedure for determining penalty
weights for constrained problems solved by approximate solvers
~\cite{Alessandroni2026}. Doucet \textit{et al.} studied how QUBO
encodings and penalty values affect the computational and thermodynamic
behavior of quantum annealers~\cite{Doucet2026}. Data-driven penalty
calibration has also been considered in application-specific settings,
including constrained molecular docking~\cite{MandalMukherjee2026}.
Other recent studies have derived problem-specific penalty bounds
~\cite{BroesamleNickel2026} or proposed alternative penalty formulations
that avoid conventional quadratic penalties and slack variables
~\cite{LeeLau2026}.
These studies confirm the importance of systematic penalty selection, but
they address different optimization problems, constraint structures,
application domains, and solver settings. A direct comparison based only
on their reported penalty values or performance results would therefore
not be informative. A fair numerical comparison would require applying
the methods to the same MBP instances, using the same QUBO formulation
and solver configuration. We position the proposed method through its methodological differences.

Hence, the contribution of this work is not the general idea of automatic penalty tuning, but an MBP-specific framework for estimating an empirically effective penalty range
from basic properties of the input graph. We first calculate a graph-dependent initial penalty estimate and then use two Gradient Boosting Regressor (GBR) models to predict the lower and upper limits of an effective multiplier interval from the number of
nodes, graph density, and the initial estimate. A final penalty value
is selected from the predicted interval and used to construct the QUBO
model. The more expensive search over candidate penalty values is thus
carried out during the offline calibration and training stage. For a
previously unseen graph, only the three input features need to be
calculated and passed to the trained models.

For the experiments, we used Erd\H{o}s--R\'enyi graphs for model calibration
and training, followed by evaluation on an independently generated test set.
These graphs were selected because their size and density can be controlled
systematically, making it possible to study how these properties influence the
penalty parameter under reproducible conditions. The resulting partitions were
compared with those produced by Metis and Kernighan--Lin in terms of partition
balance, cut size, and computational time.

The main contributions of this work are as follows:
\begin{itemize}
    \item We develop an MBP-specific penalty-tuning framework that combines a graph-dependent initial estimate with GBR to predict an effective penalty-multiplier range from basic graph properties.

    \item We build a calibration dataset of 607 Erd\H{o}s--R\'enyi graphs and evaluate the trained models on 126 independently generated instances, including hybrid-solver experiments on graphs with up to $4000$ nodes and complementary direct-QPU tests on smaller graphs with up to $100$ nodes.

    \item The learned penalty selection produced balanced partitions for all instances in the independent evaluation set and, under the adopted experimental setup, achieved lower cut values than the selected classical graph-partitioning baseline for every evaluated instance.
\end{itemize}

The remainder of this paper is organized as follows. Section~\ref{sec:Problem Definition} formally defines the MBP and presents its QUBO formulation. Section~\ref{sec:Proposed Approach} describes the experimental workflow and data-collection process. Section~\ref{sec:Penalty Parameter Calculation} introduces the graph-dependent initial penalty estimate and the empirical scaling procedure. Section~\ref{sec:machine learning--based Parameter Tuning} describes the machine learning models used for penalty prediction. Section~\ref{sec:Simulation Results} presents the hybrid-solver and direct-QPU results. Section~\ref{sec:Limitations and Future Directions} discusses possible applications, limitations, and future work. Finally, Section~\ref{sec:Conclusion} concludes the paper.

\section{Problem Definition and QUBO Formulation}
\label{sec:Problem Definition}

The MBP is formally defined as follows. Given an unweighted, undirected graph $G(V, E)$, where $V$ is the set of nodes with $|V| = n$ and $E$ is the set of edges, the objective is to partition $V$ into two equal-sized disjoint subsets\footnote{In this study, we focus only on graphs with an even number of nodes.} \(S_0, S_1 \subset V\), \(S_0 \cap S_1 =\emptyset\),  \(S_0 \cup S_1 = V\), such that $|S_0| = |S_1|$, while minimizing the number of cut edges, the edges connecting nodes from \(S_0\) to nodes from \(S_1\)~\cite{Karpinski2002}. 
The main notation used in the formulation and subsequent analysis is
summarized in Table~\ref{tab:notation}.

\begin{table}[h]
    \caption{Main notation used in the MBP formulation, penalty-parameter derivation, and machine learning analysis.}
    \label{tab:notation}
    \centering
    \renewcommand{\arraystretch}{1.2}
    \setlength{\tabcolsep}{7pt}
    \scriptsize
    \begin{tabularx}{\linewidth}{
        |>{\centering\arraybackslash}p{0.24\linewidth}|
        >{\RaggedRight\arraybackslash}X|}
        \hline
        \textbf{Symbol} & \textbf{Meaning} \\
        \hline

        \(G(V, E)\) &
        Undirected input graph with node set \(V\) and edge set \(E\). \\
        \hline

        \(n=|V|\) &
        Number of nodes in the graph; \(n\) is assumed to be even. \\
        \hline

        \(p\) &
        Edge probability used to generate an Erd\H{o}s--R\'enyi graph
        \(G(n,p)\). \\
        \hline

        \(\rho\) &
        Realized graph density,
        \(\rho=2|E|/[n(n-1)]\). \\
        \hline

        $\Delta(G)$ &
        Maximum node degree,
        $\Delta(G)=\max_{v\in V}\deg(v)$, where $\deg(v)$ denotes
        the degree of node $v$. \\
        \hline

        \(S_0,S_1\) &
        Two disjoint subsets forming the graph bisection. \\
        \hline

        \(x_i\) &
        Binary variable assigning node \(i\) to \(S_0\) or \(S_1\). \\
        \hline

        \(\mathbf{x}\) &
        Vector of binary partition variables. \\
        \hline

        \(\mathbf{Q}_{G,\lambda}\) &
        Upper-triangular QUBO matrix constructed for graph \(G\) and
        penalty parameter \(\lambda\). \\
        \hline

        \(\mathrm{cut}_{\mathrm{max}}\) &
        Largest cut of a balanced bisection, \(n^{2}/4\). \\
        \hline

        \(\mathrm{cut}(p)\) &
        Expected cut of a balanced partition in \(G(n,p)\). \\
        \hline

        \(E_{\mathrm{cut}}(\mathbf{x})\) &
        Objective term representing the number of cut edges. \\
        \hline

        \(E_{\mathrm{balance}}(\mathbf{x})\) &
        Penalty term measuring deviation from equal partition sizes. \\
        \hline

        \(E_{\mathrm{MBP}}(\mathbf{x})\) &
        Complete penalized objective function for the MBP. \\
        \hline

        \(\lambda\) &
        Final penalty coefficient enforcing equal partition sizes. \\
        \hline

        \(\lambda_{\mathrm{init}}\) &
        Objective-based initial penalty estimate from the expected cut of a balanced partition. \\
        \hline

        \(\lambda_{\mathrm{est}}\) &
        Graph-dependent initial estimate of the penalty coefficient. \\
        \hline

        \(\lambda_{\mathrm{mult}}\) &
        Multiplier applied to \(\lambda_{\mathrm{est}}\). \\
        \hline

        \(\lambda_{\min},\lambda_{\max}\) &
        Lower and upper endpoints of the effective multiplier interval. \\
        \hline

        \(\mathit{f}\) &
        Feature vector
        \((n,\rho,\lambda_{\mathrm{est}})\) used by the GBR models. \\
        \hline

        \(gbr_{\min},gbr_{\max}\) &
        GBR models predicting \(\lambda_{\min}\) and
        \(\lambda_{\max}\), respectively. \\
        \hline

        $R^{2}$, MAE, RMSE &
        Coefficient of determination, mean absolute error, and root mean square
        error; metrics used to evaluate the regression models. \\
        \hline
    \end{tabularx}
\end{table}

The rapid development of quantum computing has increased the need to reformulate
classical optimization problems so that they can be processed by quantum
hardware. To solve an optimization problem using a quantum annealer, such as
those developed by D-Wave Systems, the problem must be expressed as an Ising
model or, equivalently, in the form of a QUBO model~\cite{Glover2019,Johnson2011}.

The Ising model originates from statistical mechanics and describes a
system of interacting spins, each of which takes a value of either
\(-1\) or \(+1\). In optimization, the Ising problem consists of finding
a spin configuration that minimizes the energy of the system. Many
combinatorial optimization problems can be mapped to this
formulation~\cite{Barahona1982}. This has motivated the development of
Ising machines, which search for low-energy spin configurations using
physical or algorithmic processes such as annealing.

As with other heuristic optimization methods, their performance may be affected by
complex energy landscapes and the presence of local minima~\cite{Leleu2019}. Quantum annealers provide a different search mechanism
by using quantum fluctuations, including tunneling effects, to explore
the energy landscape~\cite{Kadowaki1998,Farhi2001,Albash2018}. These effects may improve the exploration of the solution space and help the annealer find high-quality solutions, including the global minimum in some cases.
The energy function for the Ising model is given by the Hamiltonian

\begin{equation}
\label{eq:ising}
H(\mathbf{s})
=
\sum_i h_i s_i
+
\sum_{i<j}J_{ij}s_i s_j\;,
\end{equation}

where \(\mathbf{s}\in\{-1,+1\}^{n}\) is a vector of \(n\) spin
variables, \(h_i\) is the linear bias associated with spin \(s_i\),
\(J_{ij}\) is the coupling between spins \(s_i\) and \(s_j\), and
\(H(\mathbf{s})\) is the energy of the corresponding spin configuration.

The QUBO formulation~\cite{Glover2019} is mathematically equivalent to the Ising model but instead of spin variables, it uses binary variables \( x_i \in \{0,1\}, i = 1,\dots, n \), making it more suitable for defining combinatorial optimization problems. The objective function to be minimized is

\begin{equation}
\label{qubo}
E(\mathbf{x}) = \sum_{i} \mathbf{Q}[i,i] x_i + \sum_{i<j} \mathbf{Q}[i,j] x_i x_j\;,
\end{equation}

where \( \mathbf{Q} \) is a matrix\footnote{Throughout this work, matrix \(\mathbf{Q}\) is represented in upper-triangular
form; therefore, each quadratic interaction \(x_i x_j\) with \(i<j\)
is stord and counted only once.} encoding the weights of the objective function and the penalty term and \( E(\mathbf{x}) \) represents the overall function to be minimized. The vector of \(n\) binary variables \(\mathbf{x}^{*} \in \{0,1\}^n\) is the solution to the given optimization problem if $\mathbf{x}^{*}
\in \arg\min_{\mathbf{x}\in\{0,1\}^{n}}E(\mathbf{x})$. 

Since quantum annealers natively operate in the Ising model framework~\cite{Mohseni2022, Johnson2011} and there is a linear transformation between spins and binary variables (\(x_i = \frac{s_i +1}{2}\)), the problem can be expressed in QUBO form and directly mapped onto their hardware. This compatibility allows problems such as MBP to be efficiently embedded~\cite{Dwave_minor_embedding} and solved using quantum annealers~\cite{Glover2019}.\\

The QUBO formulation of the MBP consists of two terms:
\begin{itemize}
    \item \(E_{\text{cut}}(\mathbf{x})\) --- the objective term representing the number of cut edges;
    
    \item \(E_{\text{balance}}(\mathbf{x})\) --- the penalty term that
    enforces equal partition sizes.
\end{itemize}

Based on the formulation used in related studies~\cite{Dwave_graph_partitioning}, the objective function is given by

\begin{equation}
\label{obj_function}
 E_{\text{cut}}(\mathbf{x}) = \sum_{(i,j) \in E} (x_i + x_j - 2 x_i x_j)\;,
\end{equation}

where \( x_i \in \{0,1\}, i = 1,\dots, n \), represents the binary partition assignment of node \( i \) either to set $S_1$ (in that case $x_i = 1$) or to the set $S_0$ ($x_i = 0$). The penalty term ensures that the partitions remain balanced by

\begin{equation}
\label{penalty_term}
 E_{\text{balance}}(\mathbf{x}) = \lambda \left( \sum_{i \in V} x_i - \frac{n}{2} \right)^2\;,
\end{equation}

where the penalty parameter \( \lambda \) plays a crucial role in QUBO formulation by determining the trade-off between minimizing the objective function and satisfying the penalty term.
For heuristic and quantum annealing solvers, a well-chosen
\(\lambda\) balances constraint enforcement and objective optimization.
If \(\lambda\) is too small, the penalty may be insufficient to enforce
equal partition sizes, resulting in infeasible solutions. Conversely, if $\lambda$ is set too large, the penalty term may outweigh the original objective, making it harder for an approximate solver to distinguish between solutions and potentially resulting in lower-quality partitions~\cite{Quinton2024}.

To obtain an optimal solution to the MBP, the function
$E_{\text{MBP}}(\mathbf{x})$, defined as the sum of
Eq.~\eqref{obj_function} and Eq.~\eqref{penalty_term}, must be minimized

\begin{equation}
\label{min_function}
E_{\text{MBP}}(\mathbf{x}) =  \sum_{(i,j) \in E} (x_i + x_j - 2 x_i x_j) 
 + \lambda \left( \sum_{i \in V} x_i - \frac{n}{2} \right)^2,
\end{equation}

after expansion of the quadratic term we have
\begin{align}
\label{min_function_expand}
E_{\text{MBP}}(\mathbf{x}) = & \sum_{(i,j) \in E} (x_i + x_j - 2 x_i x_j)  + \notag \\
& + \lambda \left( (1 - n) \sum_{i \in V} x_i + 2 \sum_{i < j} x_i x_j + \frac{n^2}{4} \right)\;.
\end{align}

Eq.~\eqref{min_function_expand} makes the role of the penalty parameter
explicit: the balance penalty contributes $\lambda(1-n)$ to the linear
coefficient of each variable, adds $2\lambda$ to every pairwise interaction,
and introduces the constant offset $\lambda n^2/4$, which does not affect the
location of the minimum.

The input to the quantum annealer is the matrix \(\mathbf{Q}\) of coefficients, as described in the general QUBO formulation in Eq.~\eqref{qubo}. Following~\cite{Dwave_graph_partitioning} and Eq.~\eqref{min_function_expand}, the matrix for the MBP problem is constructed as shown in Algorithm~\ref{alg:matrix}. 
Since the matrix depends on the structure of graph \(G(V, E)\) (e.g. \(\left| V \right| = \left| \mathbf{x} \right| = n \) is the order of \(\mathbf{Q}\) and matrix elements \(\mathbf{Q}[i,j]\) correspond to the coefficients of \(x_i x_j\)) and also on the penalty parameter \(\lambda\), we will denote it as \(\mathbf{Q}_{G,\lambda}\).

\begin{algorithm}[h]
    \caption{Constructing the matrix \(\mathbf{Q}_{G,\lambda}\) for QUBO formulation of MBP}
    \label{alg:matrix}
    \small
    \begin{algorithmic}
    \State 1: Initialize \(\mathbf{Q}_{G,\lambda}\) as a zero matrix
    \State 2: Update \(\mathbf{Q}_{G,\lambda}\) with coefficients from \(E_\text{cut}(\mathbf{x})\)
    \For{each edge \((i, j)\) in \(E\)}
        \State \(\mathbf{Q}_{G,\lambda}[i, i] \gets \mathbf{Q}_{G,\lambda}[i, i] + 1\)
        \State \(\mathbf{Q}_{G,\lambda}[j, j] \gets \mathbf{Q}_{G,\lambda}[j, j] + 1\)
        \State \(\mathbf{Q}_{G,\lambda}[i, j] \gets \mathbf{Q}_{G,\lambda}[i, j] - 2\)
    \EndFor
    \State 3: Update \(\mathbf{Q}_{G,\lambda}\) with coefficients from \(E_\text{balance}(\mathbf{x})\)
    \For{each node \(i\) in \(V\)}
        \State \(\mathbf{Q}_{G,\lambda}[i, i] \gets \mathbf{Q}_{G,\lambda}[i, i] + \lambda (1 - n)\)
    \EndFor
    \For{each pair of nodes \((i, j), i < j\)}
        \State \(\mathbf{Q}_{G,\lambda}[i, j] \gets \mathbf{Q}_{G,\lambda}[i, j] + 2 \lambda\)
    \EndFor
    \end{algorithmic}
\end{algorithm}

Computing $\mathbf{Q}_{G,\lambda}$ can become expensive for large graphs, both in terms of computation time and memory. The cut term is relatively straightforward to construct from a sparse adjacency matrix: the diagonal entries are obtained from the node degrees, while the quadratic terms correspond to the nonzero adjacency entries. The balance term is more demanding because it connects every pair of nodes, which leads to $\mathcal{O}(n^2)$ quadratic coefficients. Since this part has a regular structure, the coefficients can be generated more efficiently using vectorized or block-wise operations instead of processing each pair separately. Independent blocks can also be handled in parallel to reduce construction time. However, the memory requirement remains quadratic when the full binary quadratic model must be submitted to the solver.\\

\noindent
Selecting an appropriate value for \(\lambda\) is particularly challenging, as there is no \textit{universal} method for determining the optimal value. The tuning process is highly problem-specific and often relies on trial-and-error or heuristic approaches. One common and straightforward way to estimate the penalty parameter is to base it on an upper bound of the objective function~\cite{Lucas2014}. We use this approach as a starting point and first evaluate an objective-based initial estimate.

\section{Proposed Approach and Experimental Setup}
\label{sec:Proposed Approach}

The experimental procedure was designed to investigate whether suitable
QUBO penalty parameters can be estimated from measurable graph properties.
The experiments included controlled graph generation, QUBO construction,
solution using quantum, hybrid quantum-classical, and classical methods, evaluation of the resulting
partitions, and storage of the results for subsequent penalty refinement
and machine learning analysis.

\subsection{Graph Instances}
\label{subsec:graph_instances}

The MBP was evaluated on graphs generated according to the
Erd\H{o}s--R\'enyi \(G(n,p)\) model, where \(n\)\footnote{ $10 \le n \le 4000$ for the hybrid-solver experiments and
  $10 \le n \le 100$ for the direct-QPU experiments;} 
denotes the number of nodes and \(p\)\footnote{
\(p \in \{0.1,0.25,0.5,0.75\}\)}
is the probability that an edge exists between any pair of nodes.

The Erd\H{o}s--R\'enyi model was selected because graph size and expected
density can be varied systematically. This controlled setting provides
a suitable basis for examining the relationship between graph properties
and the penalty parameter and for generating consistent training and
evaluation datasets. The selected values of \(p\) cover sparse to dense
connectivity regimes within the investigated graph family.

\subsection{Solver Configuration and Comparison Scope}
\label{subsec:solver_configuration}

Several classical and quantum approaches can be used to solve graph
partitioning problems or their QUBO formulations. Table~\ref{tab:methods_comparison} summarizes their main advantages and
limitations.

\begin{table}[h]
    \caption{Main advantages and limitations of selected graph-partitioning
    and QUBO optimization approaches.}
    \label{tab:methods_comparison}
    \centering
    \renewcommand{\arraystretch}{1.2}
    \setlength{\tabcolsep}{7pt}
    \footnotesize
    \begin{tabularx}{\linewidth}{
        |>{\RaggedRight\arraybackslash}p{0.24\linewidth}|
        >{\RaggedRight\arraybackslash}X|
        >{\RaggedRight\arraybackslash}X|}
        \hline
        \multicolumn{1}{|c|}{\textbf{Approach}} &
        \multicolumn{1}{c|}{\textbf{Advantages}} &
        \multicolumn{1}{c|}{\textbf{Limitations}} \\
        \hline

        Exact MILP/MIQP solvers &
        Can provide optimality guarantees and reference solutions for
        smaller instances. &
        Computational and memory costs increase rapidly with problem size. \\
        \hline

        Multilevel partitioning, e.g., Metis~\cite{Karypis1998} &
        Fast and scalable for large graph-partitioning problems. &
        No optimality guarantee; exact balance may require postprocessing. \\
        \hline

        Local refinement, e.g., Kernighan--Lin~\cite{Kernighan1970} &
        Simple method that can preserve balanced partitions. &
        Sensitive to initialization and may converge to a local optimum. \\
        \hline

        Classical QUBO heuristics, e.g., tabu search and simulated
        annealing &
        Optimize the QUBO directly without requiring hardware embedding. &
        No optimality guarantee; performance depends on solver settings
        and runtime. \\
        \hline

        Direct-QPU quantum annealing~\cite{Albash2018} &
        Processes Ising and QUBO models and samples low-energy solutions. &
        Limited by hardware size, connectivity, embedding overhead, and
        chain breaks. \\
        \hline

        Hybrid quantum--classical solvers~\cite{Raymond2023} &
        Handle larger QUBO instances using classical and quantum
        processing. &
        Solver internals are partly proprietary, and runtime includes
        cloud and classical processing. \\
        \hline
    \end{tabularx}
\end{table}

Each graph-partitioning instance was represented as a QUBO and solved
using D-Wave Systems' hybrid solver (QA HS) through the Leap cloud
platform and the \texttt{LeapHybridSampler} class in Ocean SDK version
8.1.0. No explicit \texttt{time\_limit} was supplied, so the solver used
the minimum time automatically determined for the submitted problem
size~\cite{Dwave_Hybrid}.

Direct-QPU experiments used
\texttt{EmbeddingComposite(DWaveSampler())} with ten reads. The QPU,
minor embedding, chain strength, and annealing time were not configured
manually. Ocean therefore used the available solver (Advantage Systems) and its default
settings, including automatic chain-strength calculation and the
solver's default annealing time~\cite{Dwave_timing}. Further details can be found in Subsection~\ref{subsec:QPU testing}.

Metis and Kernighan--Lin were used as classical graph-partitioning
baselines. Algorithm Metis was executed once per graph using PyMetis version
2023.1.1 with \texttt{part\_graph(2, adjacency=G)} on unweighted graphs.
The default PyMetis settings were retained, including equal target
partition weights and the default fixed seed~\cite{KloecknerPyMetis2022}.

The custom Kernighan--Lin implementation started from a randomly
generated balanced partition and preserved balance through node-pair
swaps. It terminated when no improving swap was found or after ten
iterations. One run was performed per graph without a fixed
Kernighan--Lin seed.\footnote{The source code is available in the corresponding
\href{https://github.com/rusnakrenata/qambp/}{GitHub repository};
see the Code Availability section.}

The comparison focuses on established graph-partitioning heuristics
rather than on general-purpose QUBO or integer-programming solvers such
as simulated annealing, tabu search, or Gurobi. The objective is not to
establish the computational superiority of quantum annealing over
classical optimization, but to evaluate whether machine learning--based
penalty tuning improves the reliability and solution quality of the MBP
QUBO formulation solved by a quantum annealer. Metis and Kernighan--Lin were selected because they
operate directly on the graph-partitioning problem and provide practical
classical reference solutions.

\subsection{Experimental Workflow and Evaluation Criteria}
\label{subsec:experimental_workflow}

\begin{figure}[htbp]
    \centering
    \includegraphics[width=1.0\linewidth]{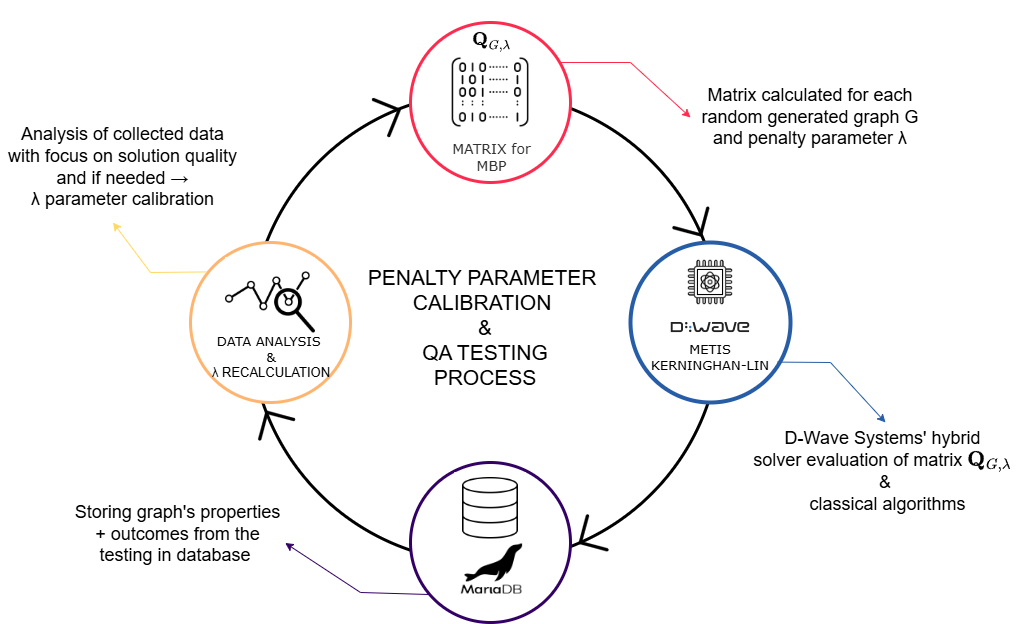}
    \caption{Workflow for penalty-parameter estimation, evaluation, and
    refinement.}
    \label{fig:process_diagram}
\end{figure}

The proposed workflow (Fig.~\ref{fig:process_diagram}) consists of the following steps

\begin{enumerate}
    \item Generate a graph \(G(n,p)\) and construct the corresponding
    QUBO matrix \(\mathbf{Q}_{G,\lambda}\) for a selected penalty
    parameter \(\lambda\).

    \item Solve the QUBO using D-Wave Systems' hybrid solver and solve
    the corresponding graph-partitioning instance using Metis and
    Kernighan--Lin algorithms.

    \item Store the graph properties, solver settings, and resulting
    partitions in a database.

    \item Evaluate solution validity and cut size, and adjust the
penalty parameter when necessary.
\end{enumerate}

The following criteria are used to evaluate and compare the methods:

\begin{itemize}
    \item \textbf{Solution validity (feasibility).} A returned assignment is
    considered a \emph{valid solution} when it satisfies the balance constraint
    exactly, i.e., $|S_0| = |S_1| = \tfrac{n}{2}$. For each method we report the
    percentage of instances for which a valid solution was found. Because Metis
    does not enforce equal partition sizes, we additionally report how often its
    partitions are balanced.

    \item \textbf{Cut size.} The primary quality measure is the number of cut
    edges $E_{\text{cut}}(\mathbf{x})$ (Eq.~\eqref{obj_function}); smaller values
    indicate better partitions. 

    \item \textbf{Relative performance.} We report the proportion of instances
    for which QA~HS achieves a lower cut than the classical alternative.

    \item \textbf{Computational time.} We measure the total runtime of each
    method; for QA~HS, the QUBO-matrix construction time is reported separately
    from the hybrid-solver time, as the two are executed on different platforms.
\end{itemize}

The statistical significance of the observed cut-size differences is assessed
later using a Wilcoxon signed-rank test (Section~\ref{sec:Simulation Results}).

\subsection{Initial Penalty Estimate}
\label{subsec:initial_penalty}

The workflow begins by constructing the QUBO matrix
\(\mathbf{Q}_{G,\lambda}\) according to Algorithm~\ref{alg:matrix}.
This construction requires an initial estimate of the penalty parameter
\(\lambda\).

For a complete graph with \(n\) nodes, a balanced partition separates
\(n^2/4\) pairs of nodes into different subsets. The corresponding cut, which
is the largest attainable for a balanced bisection, is therefore

\begin{equation}
\label{max_cut}
\mathrm{cut}_{\max} = \frac{n^2}{4}.
\end{equation}

For an Erd\H{o}s--R\'enyi graph \(G(n,p)\), each possible crossing edge is present
independently with probability \(p\). Hence, the expected cut of a fixed balanced
partition is

\begin{equation}
\label{penalty_par_max_cut}
\lambda_{\mathrm{init}} := \mathrm{cut}(p) = p\,\frac{n^2}{4}\;.
\end{equation}

Following the heuristic used in~\cite{Dwave_graph_partitioning}, this
quantity was adopted as the initial penalty estimate $\lambda_{\mathrm{init}}$.

Although $\lambda_{\mathrm{init}}$ accounts for the expected graph density
through $p$, it does not capture the structure of an individual graph
instance. Consequently, it may overestimate the penalty required to enforce
balance. An excessively large value of $\lambda$ can cause the balance term
to dominate the QUBO scale and reduce the solver's sensitivity to differences
in the cut objective~\cite{Verma2022}.

This limitation was observed in preliminary experiments on 68 graph instances
with $n \leq 4000$ and different values of $p$. The initial estimate
produced a valid QA~HS solution for $77.93\%$ of the instances. Among the
evaluated instances, the valid QA~HS solution contained fewer cut edges than
the selected classical baselines in $72.71\%$ of cases.

These results show that the density-based estimate provides a useful starting
point but is not sufficiently reliable across individual graph instances.
The following sections therefore refine it using graph-dependent information,
empirical scaling, and machine learning models that predict suitable penalty
values from graph properties.

\section{Penalty Parameter Calculation}
\label{sec:Penalty Parameter Calculation}

To refine the estimation of penalty parameter \( \lambda \), we analysed how the objective and penalty terms change in the worst case when a node is moved from one subset to the other.
 
Although the Erdős--Rényi parameter \(p\) is used in the experimental design to generate graphs with controlled density, it is intentionally excluded from the refined analytical estimation of the penalty parameter. This design choice reflects practical engineering scenarios, where the underlying graph structure is given but the generative process and its parameters are unknown.

In real-world applications, only the realized graph \(G(V, E)\) is available, while parameters such as edge probability are not directly observable. Therefore, the refined penalty estimation is formulated solely in terms of measurable graph quantities derived from \(G\), ensuring that the proposed method remains applicable to arbitrary input graphs beyond the synthetic setting.

In this context, the Erdős--Rényi model serves as a controlled data generator for training and evaluation, whereas the penalty estimation itself is designed to be model-agnostic.\\

\noindent
To establish a lower bound for $\lambda$, we assumed a complete graph and considered the case where all nodes initially belong to $ S_1 $ ($ x_i = 1, \forall i \in V) $. The objective function in this case is

\begin{equation}
    E_{\text{cut}}(\mathbf{x}) = 0\;,
\end{equation}

and the penalty term is

\begin{equation}
    E_{\text{balance}}(\mathbf{x}) = \lambda  \left( \frac{n}{2} \right)^2\;.
\end{equation}

If one node is moved from $S_1$ to $S_0$, it becomes separated from all
remaining $n-1$ nodes, and the cut objective becomes

\begin{equation}
    E_{\text{cut}}(\mathbf{x}) = n - 1\;,
\end{equation}

and the penalty term changes to

\begin{equation}
    E_{\text{balance}}(\mathbf{x}) = \lambda  \left( \frac{n}{2} - 1 \right)^2.
\end{equation}

More generally, let \(k\) denote the number of nodes moved from
\(S_1\) to \(S_0\). The two subsets then contain \(k\) and \(n-k\)
nodes, respectively. In a complete graph, every node in one subset
is connected to every node in the other subset, and the cut objective
is therefore

\begin{equation}
\label{H_cut_k}
    E_{\text{cut}}(\mathbf{x})=k(n-k)\;.
\end{equation}

The corresponding balance penalty is

\begin{equation}
\label{H_balance_k}
    E_{\text{balance}}(\mathbf{x})
    =
    \lambda\left(\frac{n}{2}-k\right)^2\;.
\end{equation}

The total energy after moving \(k\) nodes is

\begin{equation}
\label{H_total_k}
E_{\text{MBP}}(\mathbf{x})
=
k(n-k)
+
\lambda\left(\frac{n}{2}-k\right)^2\;.
\end{equation}

To ensure that an unbalanced partition does not have lower energy than
the perfectly balanced partition as in Eq.~\eqref{max_cut}, we require

\begin{equation}
E_{\text{MBP}}(\mathbf{x})
\geq \frac{n^2}{4}\;.
\end{equation}

Substituting Eq.~\eqref{H_total_k} into this condition gives

\begin{equation}
k(n-k)
+
\lambda\left(\frac{n}{2}-k\right)^2
\geq
\frac{n^2}{4}\;.
\end{equation}

The cut term can be rewritten as

\begin{equation}
k(n-k)
=
\frac{n^2}{4}
-
\left(\frac{n}{2}-k\right)^2\;,
\end{equation}

therefore,

\begin{equation}
\frac{n^2}{4}
-
\left(\frac{n}{2}-k\right)^2
+
\lambda\left(\frac{n}{2}-k\right)^2
\geq
\frac{n^2}{4}\;,
\end{equation}

which simplifies to

\begin{equation}
(\lambda-1)
\left(\frac{n}{2}-k\right)^2
\geq 0\;.
\end{equation}

Since the squared term is non-negative, this condition is satisfied when

\begin{equation}
\label{lower_bound}
\lambda \geq 1\;,
\end{equation}

which establishes the lower bound of the penalty parameter \( \lambda \).\\

\noindent
To establish a practical upper bound for \( \lambda \), we considered
a perfectly balanced partition, where \( |S_0|=|S_1|=n/2 \). In this
case, the balance penalty is zero.

We then consider the effect of moving one node from $S_1$ to $S_0$. This change increases
the balance penalty by

\begin{equation}
E_{\text{balance}}(\mathbf{x}')-
E_{\text{balance}}(\mathbf{x})
=
\lambda\;,
\end{equation}

where \(\mathbf{x}\) denotes the original balanced assignment and
\(\mathbf{x}'\) denotes the assignment after moving the node.

The edges connecting the moved node to the nodes remaining in
\(S_1\) become new cut edges. At most \(n/2-1\) such nodes remain in
\(S_1\), while the number of newly created cut edges cannot exceed the
maximum node degree
\(\Delta(G)=\max_{v\in V}\deg(v)\), where \(\deg(v)\) denotes the degree
of node \(v\). Therefore, the possible increase in the cut objective is
bounded by

\begin{equation}
E_{\text{cut}}(\mathbf{x}')
-
E_{\text{cut}}(\mathbf{x})
\leq
\min\left(
\Delta(G),
\frac{n}{2}-1
\right)\;.
\end{equation}

This bound corresponds to the largest possible increase in the cut
objective. Any removed cut edges would only reduce this increase and
therefore cannot violate the bound.

To ensure that the balance penalty remains comparable to the maximum
possible one-node increase in the cut objective, we define the following
practical upper bound for $\lambda$

\begin{equation}
\label{upper_bound}
\lambda
\leq
\min\left(
\Delta(G),
\frac{n}{2}-1
\right)\;.
\end{equation}

After calculating the lower bound in Eq.~\eqref{lower_bound}
and the upper bound in Eq.~\eqref{upper_bound}, we obtain the interval

\begin{equation}
    1 \leq \lambda \leq \min \left( \Delta(G), \frac{n}{2}-1 \right)\;.
\end{equation}

To determine a graph-dependent initial estimate for $ \lambda$, we took the midpoint of this interval

\begin{equation}
    \label{lambda_start}
    \lambda_\text{est} = \frac{1 + \min \left(\Delta(G), \frac{n}{2}-1 \right)}{2}\;.
\end{equation}
This midpoint provides a reasonable starting value for $\lambda$, avoiding both an insufficient penalty and an excessively strong balance constraint. In execution, $\lambda_{\text{est}}$ is obtained directly from the input graph, as only the maximum degree $\Delta(G)$ and the number of nodes $n$ are required and no solver call is involved. 

We followed the process shown in Fig.~\ref{fig:process_diagram} and used the newly established penalty parameter \(\lambda_\text{est}\) in the matrix \(\mathbf{Q}_{G,\lambda}\). However, after further testing on 75 randomly generated graphs with \(n \leq 4000\), we encountered several instances
for which no valid MBP solution was found. A valid solution was obtained
in 92.00\% of cases, and QA~HS outperformed the selected classical
methods in 90.67\% of cases. The estimated penalty parameter was not consistently well scaled across all graph instances, which affected both balance enforcement and cut
minimization. As a result, the algorithm failed to sufficiently minimize the number of cut edges, and in many cases, the balance condition was never satisfied, preventing the solver from finding a valid solution to MBP.

\begin{table}[h]
    \caption{Selected $\lambda_{\mathrm{mult}}$ values for different graph
    sizes based on the calibration experiments.}
    \label{tab:lambda_values}
    \centering
    \renewcommand{\arraystretch}{1.2}
    \setlength{\tabcolsep}{7pt}
    \footnotesize
    \begin{tabularx}{\linewidth}{
        |>{\centering\arraybackslash}p{0.30\linewidth}|
        >{\centering\arraybackslash}X|}
        \hline
        \textbf{Number of nodes $n$} &
        \textbf{Selected $\lambda_{\mathrm{mult}}$ values} \\
        \hline
        $100$--$200$ &
        $\{0.05, 0.1, 0.2, 0.4\}$ \\
        \hline
        $300$--$500$ &
        $\{0.005, 0.01, 0.03, 0.05, 0.1, 0.2\}$ \\
        \hline
        $600$--$900$ &
        $\{0.005, 0.01, 0.03, 0.05, 0.1\}$ \\
        \hline
        $1000$--$2000$ &
        $\{0.002, 0.005, 0.01, 0.03, 0.05, 0.1\}$ \\
        \hline
        $2500$--$4000$ &
        $\{0.0005, 0.001, 0.002, 0.005, 0.01, 0.03, 0.05, 0.1\}$ \\
        \hline
    \end{tabularx}
\end{table}
In the analysis phase of the process, we discovered that the penalty parameter \( \lambda\) must be even more scaled down for larger graphs. The multipliers \(\lambda_\text{mult}\) were selected as shown in Table~\ref{tab:lambda_values} and the penalty parameter was set as

\begin{equation}
    \label{final_lambda}
    \lambda = \lambda_\text{est} \cdot \lambda_\text{mult}\;.
\end{equation}

During the execution, \(\lambda_{\text{est}}\) is first calculated from the
properties of the input graph. It is then multiplied by a selected
\(\lambda_{\text{mult}}\) value from Table~\ref{tab:lambda_values}. The
resulting value of \(\lambda\) is used to construct the QUBO matrix
\(\mathbf{Q}_{G,\lambda}\), which is subsequently submitted to QA HS.

The different values of \(\lambda_{\mathrm{mult}}\) are needed because
\(\lambda_{\mathrm{est}}\) increases with the graph size. Both
$\Delta(G)$ and \(n/2-1\) grow approximately linearly with \(n\),
so \(\lambda_{\mathrm{est}}\) also increases approximately linearly.
However, the bounds in Eqs.~\eqref{lower_bound}
and~\eqref{upper_bound} are based on a conservative single-node
scenario, and their midpoint is often larger than the penalty required
in practice. Therefore, smaller multipliers are used for larger graphs.
The selected values decrease roughly in proportion to \(1/n\), which
compensates for the growth of \(\lambda_{\mathrm{est}}\) and keeps the
final penalty \(\lambda\) within a similar practical range across the
tested graph sizes.

After testing the candidate penalty values obtained by scaling
Eq.~\eqref{final_lambda} with the considered $\lambda_{\text{mult}}$ values on 607
randomly generated graphs, QA~HS produced at least one valid solution
for every graph. The best observed QA~HS result for each graph
outperformed the selected classical baselines in $98.53\%$ of the
cases.

\begin{figure}[h]
\caption{Comparison of Kernighan--Lin, PyMetis, and QA HS cut-edge
results across different numbers of nodes, $n \le 1000$. Larger sizes are omitted because the cut counts grow approximately as $\rho n^2 /4$, which compresses the smaller sizes to visually indistinguishable bars; The
axis label ``inter-edges'' denotes the number of cut edges
$E_{\mathrm{cut}}(\mathbf{x})$ from Eq.~\eqref{obj_function}.}
    \label{fig:kl_pm_qa_comparison}
    \centering
    \includegraphics[width=1.0\linewidth]{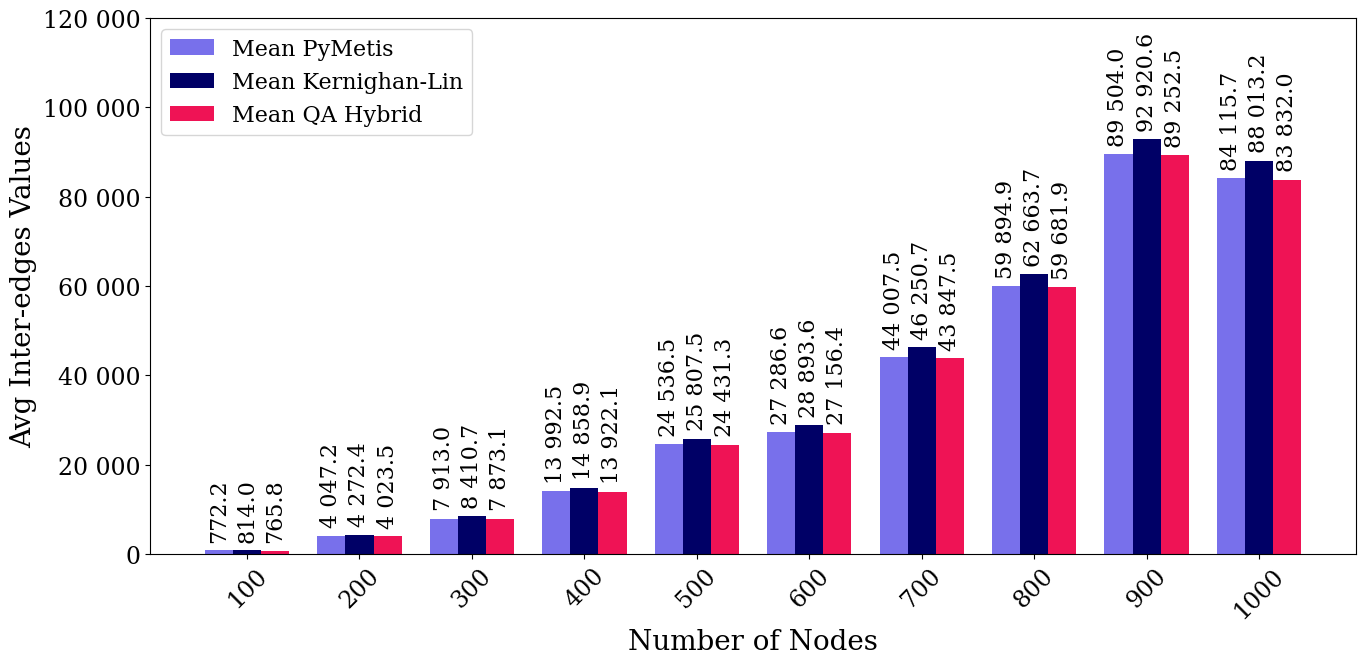}
\end{figure}

The results shown in Fig.~\ref{fig:kl_pm_qa_comparison} indicate that
the best observed QA HS solutions obtained during penalty calibration
generally produced fewer cut edges than PyMetis and Kernighan--Lin.
Kernighan--Lin produced the largest cut at every graph size tested and
therefore offered no additional insight beyond PyMetis as a classical
reference. It was retained during calibration as a second classical
reference point but excluded from the independent evaluation in
Section~\ref{sec:Simulation Results}.

Additionally, the QA HS cut edges values were selected based on the best results for the given penalty parameter multipliers from the \( \lambda_{\text{mult}} \) values. In cases where multiple \( \lambda_{\text{mult}} \) values led to the same minimal cut edges, we chose the minimum \( \lambda_{\text{min}} \) and maximum \( \lambda_{\text{max}} \) penalty-multiplier values. These values together formed a range \( \left[ \lambda_{\text{min}},  \lambda_{\text{max}} \right] \) representing an interval of tested values that produced the best observed partitioning results and were later used as target values for training the machine learning--based prediction model.

\begin{figure}[h]
\caption{Ranges of $\lambda_{\text{mult}}$ values by graph size that
produced the lowest number of cut edges with QA~HS, derived from
experiments on 607 randomly generated Erd\H{o}s--R\'enyi graphs.}
    \label{fig:lambda_ranges}
    \centering
    \includegraphics[width=1.0\linewidth]{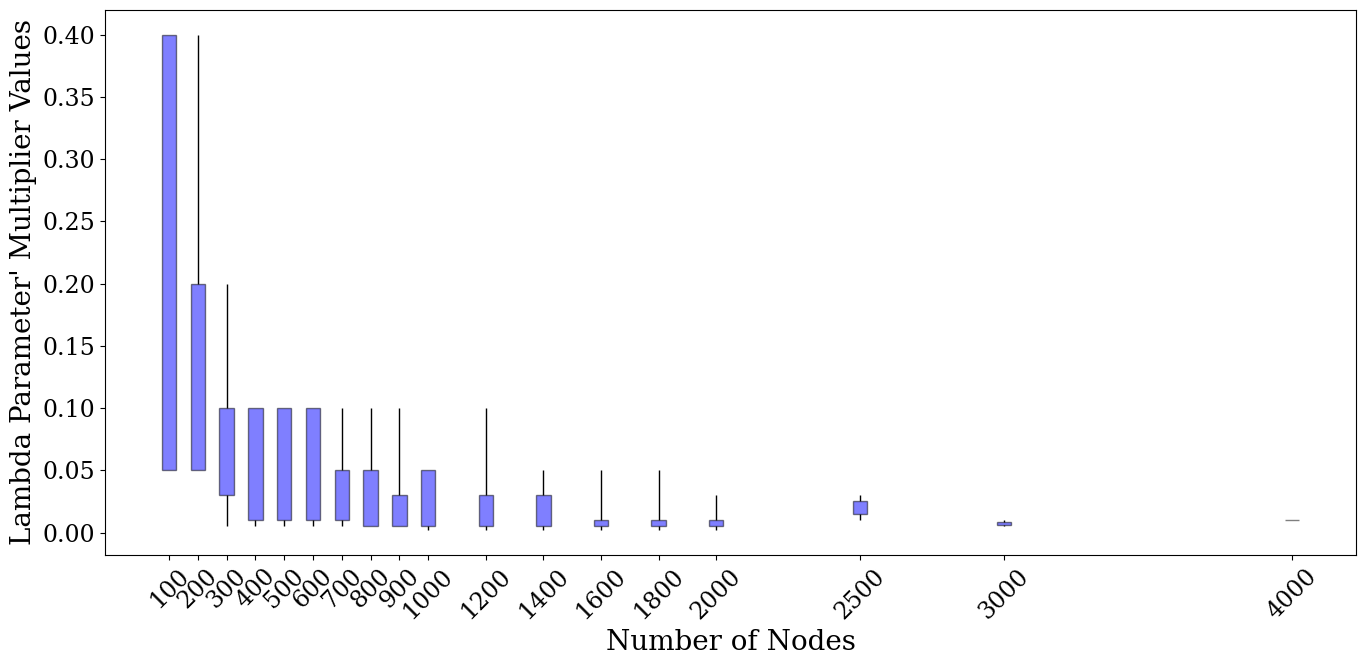}
\end{figure}

To illustrate the relationship between the number of nodes and the \( \lambda_{\text{mult}} \) parameter ranges, Fig.~\ref{fig:lambda_ranges} presents a candle plot. 
Although the obtained results were satisfactory, determining the interval \( [\lambda_{\text{min}}, \lambda_{\text{max}}] \) required executing the QA HS for all candidate values listed in Table~\ref{tab:lambda_values} and evaluating the corresponding solution quality. 
To eliminate this costly manual search, we therefore trained a machine
learning model on the collected experimental data to predict suitable
values of \(\lambda_{\mathrm{mult}}\) from the graph properties.

\section{Machine Learning--Based Parameter Tuning}
\label{sec:machine learning--based Parameter Tuning}
Initially, we considered several graph properties as independent variables, including number of nodes (\( n \)), number of edges (\( |E| \)), density (\( \rho \)), graph modularity and community structure.
Since the graph density is defined as \(\rho = \frac{2|E|}{n(n-1)}\), the number of edges is fully determined by \(n\) and \(\rho\). Therefore, including \(|E|\) as an additional feature
would introduce redundant information, and it was excluded from the
final feature set. Additionally, modularity and community structure were also considered~\cite{Fortunato2012}, but computing them would require selecting a community-detection algorithm, introducing a further preprocessing step and its own parameters into the feature definition. These properties were therefore excluded from the
final feature set.

Before implementing the GBR--based model for penalty parameter \(\lambda\) prediction, we conducted an analysis to determine how often specific penalty-multiplier values \(\lambda_\text{mult}\) from Table~\ref{tab:lambda_values} led to the minimum cut edges for MBP. The primary objective was to evaluate the success rate of the QA HS under different graph configurations and chosen \(\lambda_\text{mult}\) values, where the penalty parameter used in the underlying QUBO matrix is given as in Eq.~\eqref{final_lambda}.

\begin{figure}[h]
    \caption{Success rate of the $\lambda_{\mathrm{mult}}$ for graphs with different $\rho$ and $n$ restricted to $n\le 2000$. The color encodes the proportion of instances for which a candidate $\lambda_{\mathrm{mult}}$ gave the
smallest QA~HS cut. Larger sizes used a different multiplier set with
too few instances per combination.}
    \label{fig:success_rate_heatmap}
    \centering
    \includegraphics[width=0.95\linewidth]{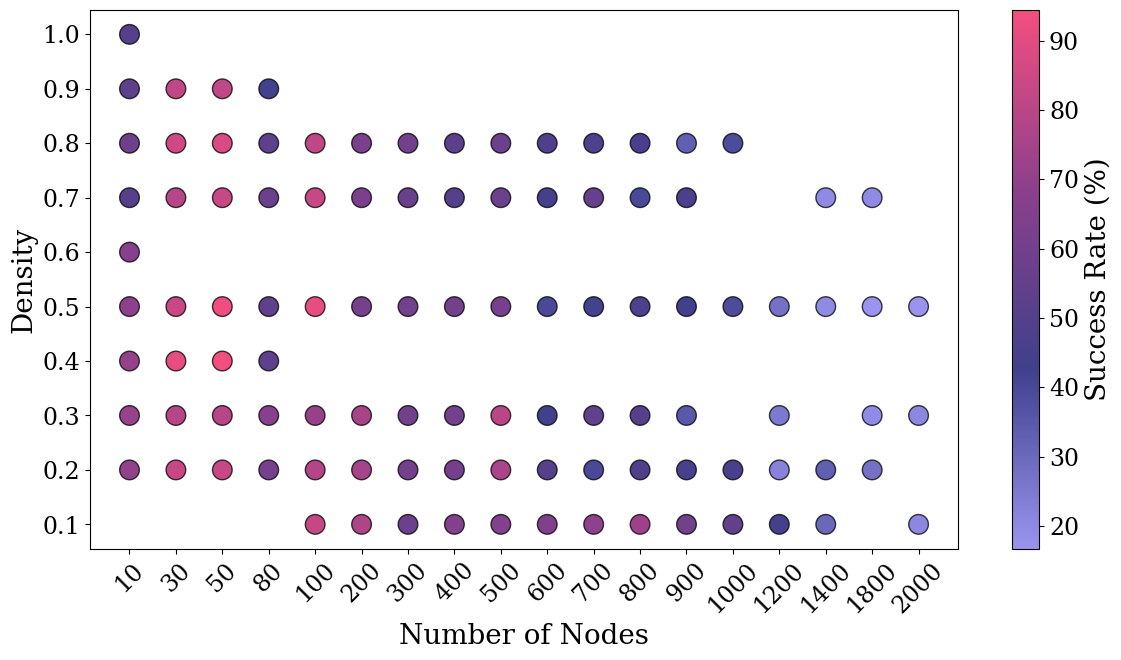}
\end{figure}

Fig.~\ref{fig:success_rate_heatmap} presents a heatmap visualization of the QA HS success rates across various numbers of nodes and graph densities. The color intensity represents the probability that, under a given choice of \(\lambda_\text{mult}\), the solver produced the best observed valid partition.
The success rate was computed as the proportion of test cases where the QA HS produced the minimum observed cut value among the tested \(\lambda_\text{mult}\) values. From the visualization, we observe the following:

\begin{itemize}
    \item The QA HS achieved higher success rates at moderate graph densities (\(0.25 - 0.75\)), whereas very sparse or very dense graphs resulted in lower success rates.
    \item Larger graphs (\(n \geq 1000\)) displayed relatively stable success rates, indicating that the solver scales well when $\lambda$ is properly chosen.
    \item The heatmap highlights the dependency of solver performance on density, suggesting that selecting lambda adaptively based on graph structure is critical.
\end{itemize}

This analysis underscores the importance of tuning solver parameters dynamically rather than relying on fixed calculated values. It also provided motivation for implementing machine learning--based penalty parameter selection. Hence, to optimize \(\lambda\) selection further, we trained two GBR, machine learning--based models, to predict the lower and upper bound of the penalty-multiplier range \( \left[ \lambda_{\text{min}},  \lambda_{\text{max}} \right] \), see Algorithm~\ref{alg:GBR_training}. 

\begin{algorithm}[h]
    \caption{Training the GBR model}
    \label{alg:GBR_training}
    \begin{algorithmic}[1]
    \small
    \State Extract collected graph properties ($n,  |E|,  \Delta(G) $) and (\( \lambda_\text{min} ,  \lambda_\text{max}\)) from database.
    \State Compute $\rho = \frac{2|E|}{n(n-1)}$ and form
       $f = (n, \rho, \lambda_{\mathrm{est}})$.
    \State Train two GBR models (\( gbr_{\text{min}}, gbr_{\text{max}} \)) to predict (\( \lambda_\text{min} , \lambda_\text{max}\)).
    \State Evaluate models on the test subset using RMSE, MAE, $\text{R}^2$ score.
    \State Save trained models for future predictions.
    \end{algorithmic}
\end{algorithm}

The two GBRs were implemented in Python using scikit-learn version 1.6.1.
Separate models were trained to predict $\lambda_{\min}$ and
$\lambda_{\max}$. The input features were the number of nodes $n$, the
realized graph density $\rho$, and the initial penalty estimate
$\lambda_{\text{est}}$. The dataset of 607 graph instances was divided
into 485 training instances and 122 test instances using an 80:20 split
with \texttt{random\_state=42}. The same random seed was used for both
regressors. No hyperparameter optimization was performed; apart from the
random seed, the default \texttt{GradientBoostingRegressor} configuration
was retained\footnote{Specifically, the GBR configuration was
\texttt{n\_estimators=100}, \texttt{learning\_rate=0.1},
\texttt{max\_depth=3}, \texttt{loss='squared\_error'},
\texttt{criterion='friedman\_mse'}, \texttt{subsample=1.0},
\texttt{min\_samples\_split=2}, \texttt{min\_samples\_leaf=1},
\texttt{max\_features=None}.} 
and early stopping was disabled by setting
\texttt{n\_iter\_no\_change=None}. The adequacy of this configuration is
assessed in Section~\ref{subsec:Hybrid Solver Testing} through the downstream evaluation on unseen graph instances. No feature scaling or normalization was applied.
Performance was evaluated on the test subset using MAE, RMSE, and $R^2$.

The GBR model~\cite{Friedman2001} was selected
because the learning task involves a relatively small tabular dataset
with 607 samples and three input features. Tree-based ensemble methods
are well suited to this type of data and can capture nonlinear
relationships without extensive feature engineering
~\cite{Grinsztajn2022,Shwartzziv2022}. Gradient boosting has also shown
strong predictive performance in related optimization and
parameter-tuning tasks~\cite{Zheng2017}.

A systematic comparison with alternative models such as Random Forest, XGBoost,
LightGBM, and neural networks was not performed. The objective of this study was
to determine whether suitable penalty multipliers could be learned from basic
graph properties and used successfully in the complete optimization workflow,
rather than to identify the best possible regression algorithm. The
implementation is modular and model-agnostic. In the publicly available source
code, the regression stage is contained in the
\texttt{gbr\_prediction.py} module, which can be replaced with another
regression model without changing the remaining stages of the workflow.

\begin{figure}[h]
\caption{Predicted versus true penalty multiplier values $\lambda_{\min}$ and $\lambda_{\max}$, obtained from the trained models $gbr_{\min}$ and $gbr_{\max}$. The dashed line represents the ideal perfect fit.}
    \label{fig:lambda_fit}
    \centering
    \includegraphics[width=1.0\linewidth]{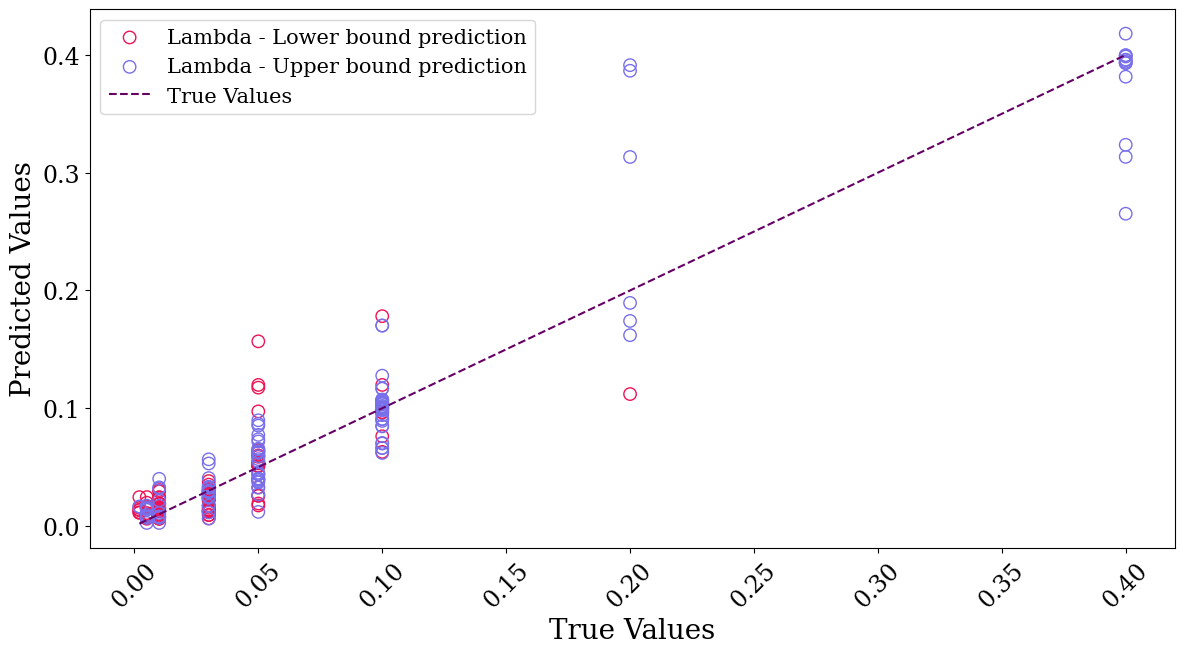} 
\end{figure}

From Fig.~\ref{fig:lambda_fit} we can see different levels of predictive performance for
the two interval bounds. For the
upper bound $\lambda_{\text{max}}$, the model achieved an $\text{R}^2$
score of 0.9028, with RMSE of 0.0360 and MAE of 0.0188, indicating strong
predictive accuracy. For the lower bound $\lambda_{\text{min}}$, the model
attained an $\text{R}^2$ score of 0.4489, with MAE of 0.0124 and RMSE of
0.0211, indicating moderate explanatory capability.

\begin{figure}[h]
    \caption{Distribution of the observed $\lambda_{\text{min}}$ and
$\lambda_{\text{max}}$ values across the graph instances used to train
the GBR models.}
    \label{fig:lambda_distribution}
    \centering
    \includegraphics[width=1.0\linewidth]{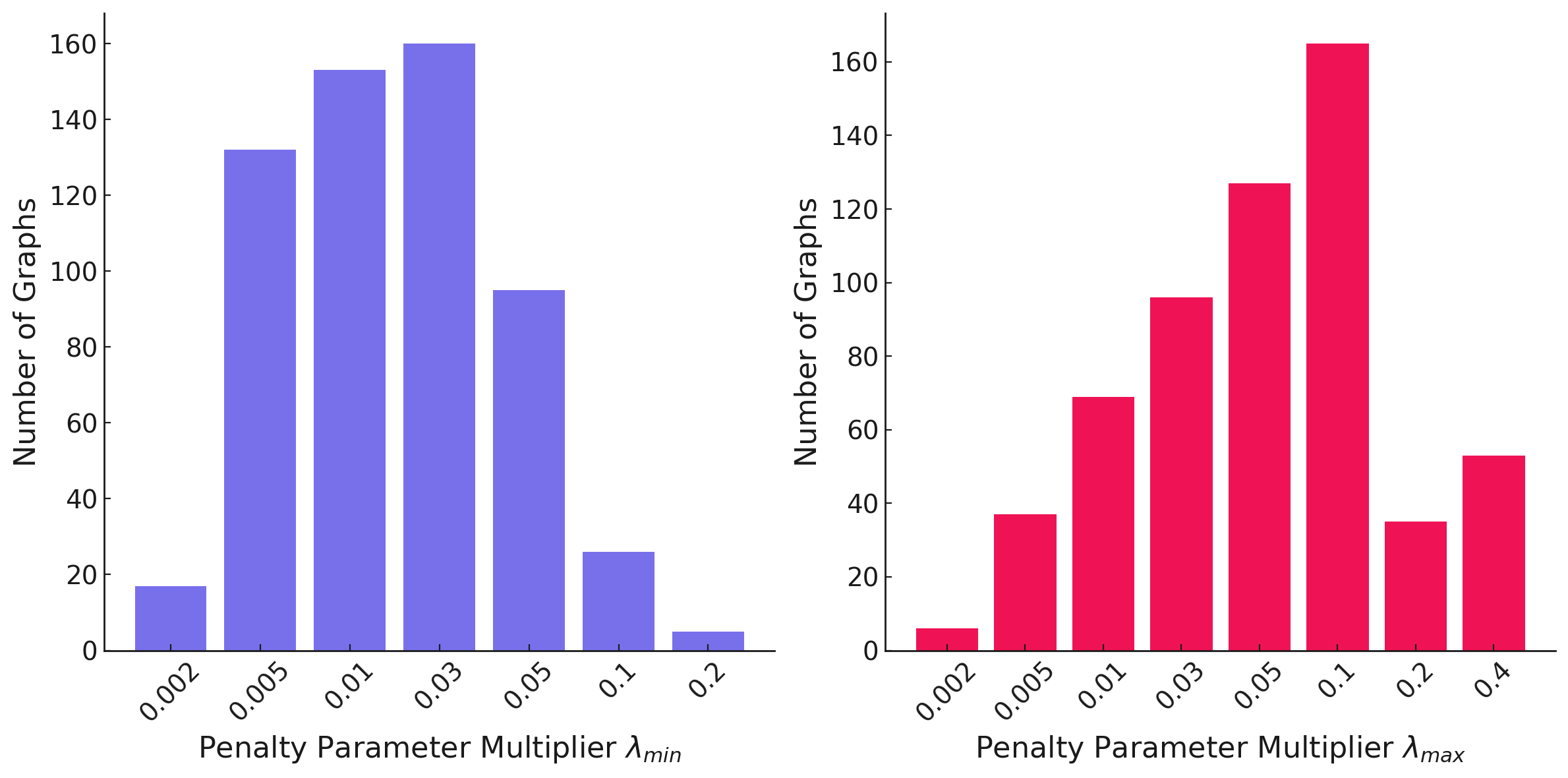}
\end{figure}

These results are consistent with the empirical distributions of
$\lambda_{\text{min}}$ and $\lambda_{\text{max}}$ across graph instances
(Fig.~\ref{fig:lambda_distribution}). The values of
$\lambda_{\text{min}}$ are highly concentrated within a narrow interval,
which inherently limits the variance available for learning, whereas
$\lambda_{\text{max}}$ exhibits a broader and more structured
distribution that is easier to model. Because the coefficient of
determination is sensitive to the variance of the target variable, the
result for $\lambda_{\text{min}}$ should also be interpreted together
with its small absolute errors. The reported RMSE and MAE indicate that
the predictions remained close to the observed multiplier values despite
the limited variation in the target.

The final predicted values for penalty parameter \(\lambda\) were obtained by combining the lower and upper bound predictions 

\begin{equation}
    \label{lambda_gbr}
   \lambda = \lambda_\text{est}  \frac{\lambda_\text{min} + \lambda_\text{max}}{2}\;,
\end{equation}
as shown in Algorithm~\ref{alg:GBR_prediction}. During execution, for a previously unseen graph the feature vector $\mathit{f}=(n,\rho,\lambda_\text{est})$ is first computed; the trained regressors $gbr_\text{min}$ and $gbr_\text{max}$ then predict $\lambda_\text{min}$ and $\lambda_\text{max}$, and the final penalty $\lambda$ follows from Eq.~\eqref{lambda_gbr} without any additional solver runs.

\begin{algorithm}[h]
    \caption{Predicting $\lambda$ for a new graph}
    \label{alg:GBR_prediction}
    \begin{algorithmic}[1]
    \small
    \State Compute $ \lambda_\text{est} = \frac{1 + \min(\Delta(G), n/2 - 1)}{2} $ and  $\Delta(G) = \max_{v \in V} \deg(v)$ for new graph $G(V, E)$ with $|V| = n$.
    \State Compute $\rho = \frac{2|E|}{n(n-1)}$ and construct feature vector $ \mathit{f} = (n, \rho, \lambda_\text{est}) $.
    \State Predict ($ \lambda_\text{min} , \lambda_\text{max} $) using trained regressors ($ gbr_\text{min}, gbr_\text{max}$).
    \State Compute final lambda as $ \lambda \gets \lambda_\text{est}  \frac{\lambda_\text{min} + \lambda_\text{max}}{2} $.
    \end{algorithmic}
\end{algorithm}

\section{Experimental Results}
\label{sec:Simulation Results}

In this section, we present the results of our simulations after GBR-based penalty parameter tuning using the QA HS for graphs with \(n \leq 4000\) nodes, as well as the QPU for graphs with \(n \leq 100\) nodes.

All computations, including graph generation, QUBO matrix construction, analytical penalty estimation, and machine learning model training, were performed classically. Quantum resources were used exclusively for solving the resulting QUBO instances via the QPU and the QA HS solver. All components of the experimental pipeline  were implemented in Python. The results and experiment metadata were stored in a MariaDB relational database for further analysis.

The classical computations were executed on a dedicated Linux VPS server equipped with an Intel Core Processor (Haswell, no TSX, IBRS), featuring 3 CPU cores running at 2.6~GHz and 6~GB of RAM, operating on Ubuntu Server 24.04 LTS (64-bit). The architectural characteristics, advantages, and limitations of the D-Wave Systems' platform are discussed in subsequent subsections.

\subsection{Hybrid Solver Testing}
\label{subsec:Hybrid Solver Testing}

To evaluate QA HS after tuning \(\lambda\) using the GBR model in
Eq.~\eqref{lambda_gbr}, the predicted \(\lambda\) values were used to
construct the QUBO matrices \(\mathbf{Q}_{G,\lambda}\) for 126 newly
generated random graphs. The resulting QA HS solutions were then compared
with those produced by PyMetis in terms of solution quality and computational time.

Additionally, we calculated the percentage of cases in which PyMetis produced a balanced solution\footnote{In the Metis algorithm, the balance constraint is not enforced.} and the percentage of cases where QA HS successfully found a valid solution, which was constantly \(100\%.\) 
Table~\ref{tab:after_regression} presents a comparative analysis of the performance of both solvers. 

\begin{table}[h]
    \caption{Comparison of QA~HS and PyMetis for the MBP after
    GBR-based penalty-parameter tuning.}
    \label{tab:after_regression}
    \centering
    \renewcommand{\arraystretch}{1.2}
    \setlength{\tabcolsep}{4pt}
    \footnotesize
    \begin{tabularx}{\linewidth}{
        |>{\hsize=0.85\hsize\centering\arraybackslash}X|
        >{\hsize=0.75\hsize\centering\arraybackslash}X|
        >{\hsize=1.25\hsize\centering\arraybackslash}X|
        >{\hsize=1.00\hsize\centering\arraybackslash}X|
        >{\hsize=1.05\hsize\centering\arraybackslash}X|
        >{\hsize=1.10\hsize\centering\arraybackslash}X|}
        \hline
        \textbf{Number of nodes \(n\)} &
        \textbf{Average density $\bar{\rho}$} &
        \textbf{PyMetis balanced (\%)} &
        \textbf{QA~HS better (\%)} &
        \textbf{Absolute difference\(^{a}\)} &
        \textbf{Percentage difference\(^{b}\)} \\
        \hline

        100  & 0.715 & 66.7  & \(\mathbf{100}\) & 6.0    & 0.361 \\
        \hline
        200  & 0.438 & 40.0  & \(\mathbf{100}\) & 19.2   & 0.485 \\
        \hline
        300  & 0.360 & 66.7  & \(\mathbf{100}\) & 58.7   & 0.808 \\
        \hline
        400  & 0.379 & 57.1  & \(\mathbf{100}\) & 62.4   & 0.446 \\
        \hline
        500  & 0.766 & 0.0   & \(\mathbf{100}\) & 94.3   & 0.204 \\
        \hline
        600  & 0.437 & 75.0  & \(\mathbf{100}\) & 141.0  & 0.380 \\
        \hline
        700  & 0.521 & 80.0  & \(\mathbf{100}\) & 217.6  & 0.360 \\
        \hline
        800  & 0.259 & 60.0  & \(\mathbf{100}\) & 180.0  & 0.471 \\
        \hline
        1000 & 0.495 & 36.4  & \(\mathbf{100}\) & 249.4  & 0.209 \\
        \hline
        1500 & 0.750 & 0.0   & \(\mathbf{100}\) & 622.0  & 0.151 \\
        \hline
        2000 & 0.600 & 0.0   & \(\mathbf{100}\) & 1243.0 & 0.213 \\
        \hline
        3000 & 0.138 & 25.0  & \(\mathbf{100}\) & 1119.3 & 0.388 \\
        \hline
        4000 & 0.100 & 100.0 & \(\mathbf{100}\) & 1372.0 & 0.369 \\
        \hline
    \end{tabularx}

    \vspace{2pt}
    \begin{minipage}{\linewidth}
        \footnotesize
        \(^{a}\) Absolute difference:
        \(\left|
        \operatorname{avg}(\mathrm{cut}_{\mathrm{QA\,HS}})
        -
        \operatorname{avg}(\mathrm{cut}_{\mathrm{PyMetis}})
        \right|\).

        \(^{b}\) Percentage difference:
        \(\dfrac{
        \operatorname{avg}(\mathrm{cut}_{\mathrm{PyMetis}})
        -
        \operatorname{avg}(\mathrm{cut}_{\mathrm{QA\,HS}})
        }{
        \operatorname{avg}(\mathrm{cut}_{\mathrm{QA\,HS}})
        }\times100\%\).
    \end{minipage}
\end{table}

From the data, we observe that:
\begin{itemize}
\item PyMetis was outperformed by QA HS in \(100\%\) of cases, indicating that QA HS produced better partitions in every tested MBP instance.
\item PyMetis provided a balanced solution in approximately \(50\%\) of cases.
\item The percentage difference shows
no monotone trend with graph size: the largest value occurs at
$n = 300$ and the smallest at $n = 1500$.
\item The average graph density varied between
$\bar{\rho}=0.100$ and $\bar{\rho}=0.766$. Despite this variability,
QA~HS consistently outperformed PyMetis across all tested graph sizes.
\item The absolute difference in cut edges increases with the number of nodes, reaching a maximum of 1372 cut edges at \( n = 4000 \).
\end{itemize}

To assess whether QA~HS and PyMetis differed systematically, a two-sided
Wilcoxon signed-rank test was applied to the paired cut-size differences,
defined as
$d_i=\mathrm{cut}_{\mathrm{PyMetis},i}
-\mathrm{cut}_{\mathrm{QA\,HS},i}$,
$i=1,\ldots,126$. The null hypothesis assumed that the paired
differences were centered at zero, whereas the alternative hypothesis
assumed a non-zero difference. All paired differences were positive,
favoring QA~HS. The test yielded $W=0$, the minimum possible value of
the statistic, and $p<0.001$. Therefore, the null hypothesis was
rejected at $\alpha=0.05$.\footnote{The test was performed using the
\texttt{scipy.stats.wilcoxon} function from SciPy version 1.15.1.} Because all 126 differences had the same direction, the $p$-value reflects the consistency of the improvement rather than its magnitude. The magnitude of the improvement is reported by the percentage
differences in Table~\ref{tab:after_regression}, which range from
approximately $0.15\%$ to $0.81\%$.

\begin{figure}[h]
\caption{Absolute and percentage differences in cut edges between QA HS
and PyMetis. ``Inter-edges'' in the axis label denotes cut edges.}
    \label{fig:after_testing_comparison}
    \centering
    \includegraphics[width=1.0\linewidth]{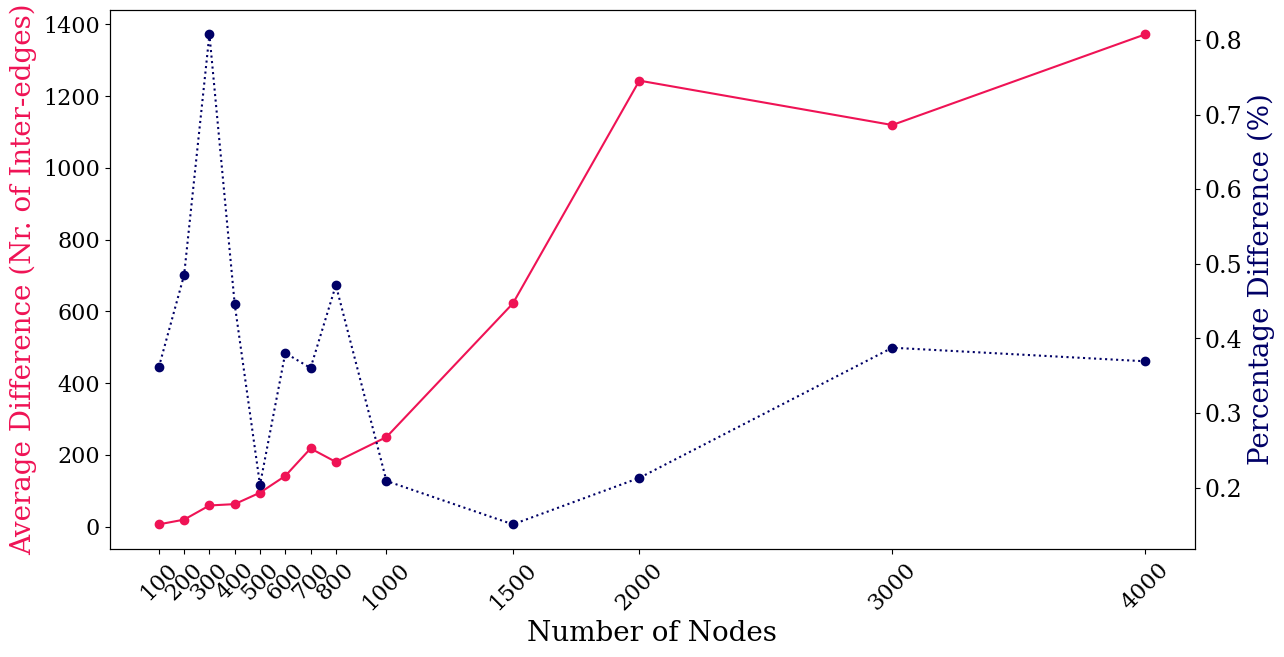}
\end{figure}

Further, Fig.~\ref{fig:after_testing_comparison} graphically represents the absolute and percentage differences in cut edges between QA HS and PyMetis. The absolute difference generally increases with the number of nodes, whereas the percentage difference fluctuates for smaller graphs and
becomes more stable for larger graphs.

This analysis indicates that combining QUBO-based modeling with data-driven penalty parameter tuning enabled QA~HS to produce lower cut values than PyMetis under the adopted experimental setup. In particular, the integration of the GBR-predicted penalty parameter allowed the quantum--classical pipeline to balance constraint enforcement and cut
minimization more effectively. The fact that all 126 independently generated instances produced valid balanced solutions and lower cut values than PyMetis indicates that the selected GBR configuration was adequate for the purpose of this study.

In addition to solution quality, we also analyzed computational time. For classical algorithms, this refers to the total execution time of the partitioning process, whereas for the QA HS, we accounted for additional factors. Because the QA HS is a cloud-based solver with hybrid processing, its runtime includes multiple components beyond annealing time, such as Binary Quadratic Model (BQM) construction, embedding, preprocessing, quantum-classical post-processing~\cite{Dwave_timing}, cloud communication with network latency, as well as the QUBO matrix calculation. This must be considered when interpreting computational efficiency, as the response time includes both quantum and classical components. Here we measured the overall computational time and QUBO matrix separately. 
The QUBO matrix calculations were executed on a classical computer. This step is computationally demanding due to the pairwise iteration required for large numbers of nodes, see Algorithm~\ref{alg:matrix}.

For graphs with \(n \geq 1000\), PyMetis was approximately three times
faster than the overall QA~HS workflow. For smaller graphs, PyMetis showed an even greater performance advantage. The QUBO matrix calculation was not optimized and grew significantly with the number of nodes, taking 163 seconds for graphs with $4000$ nodes, but the matrix calculation optimization is outside the scope of this article.\footnote{Further details on possible improvements to the matrix
construction can be found in Section~\ref{sec:Problem Definition}.} It is also important to note that its performance as well as the PyMetis execution time is highly dependent on the properties of the classical resources used.

\subsection{QPU Testing}
\label{subsec:QPU testing}

To evaluate direct-QPU execution, we conducted experiments using a D-Wave Advantage QPU accessed through the Ocean SDK. As a classical baseline, we again used PyMetis. 

Unlike the hybrid solver, whose internal decomposition and use of
quantum and classical resources are managed automatically by D-Wave's
Leap quantum cloud service, direct-QPU execution requires the submitted QUBO to be embedded explicitly into the hardware topology. Consequently, the size and density of directly executed problems are constrained by the available number of qubits, hardware connectivity, and embedding overhead~\cite{Dwave_minor_embedding}.

The D-Wave Advantage system with Pegasus topology provides 5640 qubits, each connected to up to 15 neighbors, enabling the embedding of complete graphs with up to 124 nodes using chain lengths of approximately 17~\cite{DWave_Advantage2021}. The practically embeddable size, however, also depends strongly on graph density and structure~\cite{Boothby2020}.

In this study, the problem embedding onto the QPU topology was performed automatically
using the \texttt{EmbeddingComposite} wrapper, which uses the
\texttt{minorminer} heuristic to construct a minor embedding. Logical variables
were represented by chains of physical qubits whose lengths were determined by
the embedding. Because no chain strength was supplied explicitly,
\texttt{EmbeddingComposite} calculated it automatically using the default
\texttt{uniform\_torque\_compensation} method. Broken chains were resolved using
the default majority-vote postprocessing method. All QPU runs used the
solver-default annealing time of $20~\mu$s and 10 reads per problem instance.
No custom annealing schedule, anneal offsets, flux-bias corrections, or other
hardware-level parameter tuning was applied.

\begin{figure}[h]
    \caption{Comparison of QPU and PyMetis partitioning results on graphs with $n \leq 100$.}
    \label{fig:qpu_vs_pymetis}
    \centering
    \includegraphics[width=0.9\linewidth]{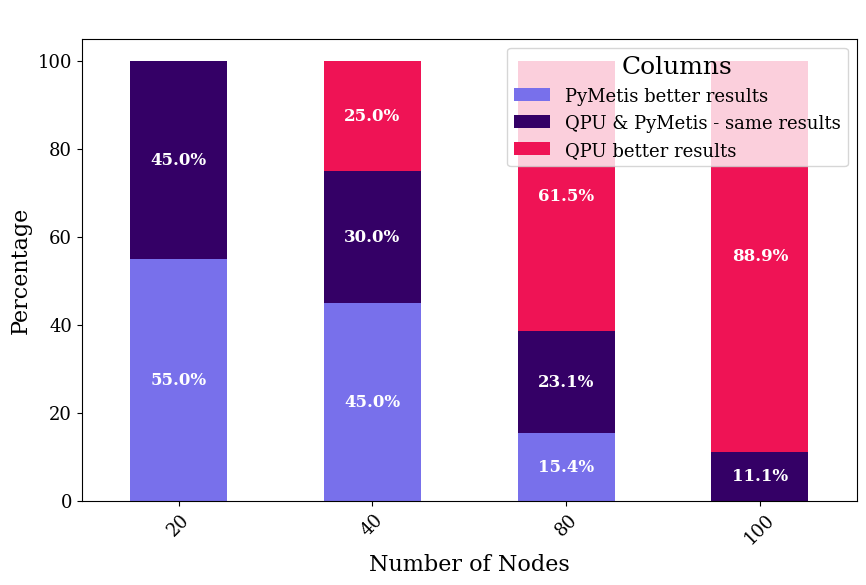}
\end{figure}

For testing, we selected 70 randomly generated Erdős--Rényi graphs with $n \leq 100$ to ensure compatibility with the QPU’s connectivity constraints.
From Fig.~\ref{fig:qpu_vs_pymetis}, the following observations can be made:
\begin{itemize}
  \item For the smallest graphs ($n = 20$ and $n = 40$), PyMetis produced
        a smaller cut more often than the QPU (55.0\% and 45.0\% of
        instances, respectively), and the QPU never produced a smaller cut
        at $n = 20$.
  \item For the larger instances ($n = 80$ and $n = 100$), the QPU more
        frequently produced partitions with fewer cut edges, in 61.5\% and
        88.9\% of instances, respectively.
  \item The two methods therefore cross over between $n = 40$ and
        $n = 80$. This behavior is partly explained by the fact that the
        penalty parameter $\lambda$ was predicted using a model trained
        primarily on graphs with $n \ge 100$, which may result in
        suboptimal parameter selection for smaller instances.
\end{itemize}

Despite these promising results for instances with $n \geq 80$, fully quantum execution remains limited by several hardware-related factors. Most notably, minor embedding introduces a substantial overhead and variability in solution quality due to chain breaks, particularly for dense graphs. 

Additional limitations arise from restricted qubit connectivity~\cite{Quinton2024}, the need for careful chain strength selection~\cite{Choi2008p1, Raymond2020}, and sensitivity to annealing schedule parameters~\cite{Galindo2020, Perdomo-Ortiz2016}. While further tuning of these parameters could improve performance, such hardware-specific optimization is outside the scope of this study.

Given these constraints, hybrid quantum--classical solvers currently represent a more practical solution for graphs with \(n \geq 100\), as they combine the scalability and robustness of classical optimization with the sampling capabilities of quantum annealing.

\subsection{Summary}
\label{subsec:Summary}

Table~\ref{tab:lambda_tuning_summary} summarizes the impact of progressively refined penalty parameter selection strategies on the performance of the hybrid quantum annealing solver. 

\begin{table}[h]
    \caption{Performance of QA~HS under progressively refined
    penalty-parameter selection strategies.}
    \label{tab:lambda_tuning_summary}
    \centering
    \renewcommand{\arraystretch}{1.2}
    \setlength{\tabcolsep}{7pt}
    \footnotesize
    \begin{tabularx}{\linewidth}{
        |>{\hsize=1.25\hsize\centering\arraybackslash}X|
        >{\hsize=0.65\hsize\centering\arraybackslash}X|
        >{\hsize=1.10\hsize\centering\arraybackslash}X|}
        \hline
        \textbf{Strategy for setting \(\lambda\)} &
        \textbf{Valid solutions (\%)} &
        \textbf{QA~HS better than classical baselines (\%)} \\
        \hline

        $\lambda=\lambda_{\mathrm{init}}=\mathrm{cut}(p)$ &
        77.93 &
        72.71 \\
        \hline

        \(\lambda=\lambda_{\mathrm{est}}\) &
        92.00 &
        90.67 \\
        \hline

        \(\lambda=
        \lambda_{\mathrm{est}} \cdot \lambda_{\mathrm{mult}}\) &
        \(\mathbf{100.00}\) &
        98.53 \\
        \hline

        \(\lambda\) predicted using GBR &
        \(\mathbf{100.00}\) &
        \(\mathbf{100.00}\) \\
        \hline
    \end{tabularx}
\end{table}

The initial estimate based on the expected cut led to a solution feasibility of only 77.93\% and limited improvement over classical methods. Incorporating analytically derived bounds increased the success rate to 92.00\% and improved solution quality in 90.67\% of cases. Introducing an additional scaling factor further stabilized the optimization process, resulting in valid solutions for all tested instances and outperforming classical heuristics in 98.53\% of cases. The best results were achieved using the machine learning–predicted penalty parameter, which consistently yielded feasible solutions and superior partitions across all evaluated graphs.

Complementary experiments on D-Wave’s QPU indicate that the predicted penalty values can also be applied to small problem instances in direct-QPU execution, although hardware connectivity and embedding constraints currently limit scalability compared with the hybrid solver.

Overall, these results indicate that systematic, machine learning–assisted penalty parameter tuning can improve the reliability and competitiveness of QUBO-based graph partitioning on the tested Erd\H{o}s--R\'enyi graphs. They also suggest that such tuning may support the more effective use of quantum and hybrid solvers in broader optimization workflows.

\section{Limitations and Future Directions}
\label{sec:Limitations and Future Directions}

The proposed penalty-tuning framework has potential applications in areas
where balanced graph partitioning is required, including transportation
networks, distributed computing, data clustering, image segmentation,
VLSI design, and social-network analysis. In such applications, balanced
partitions can support workload distribution, reduce communication
overhead, or separate large networks into smaller computational units.
However, the practical applicability of the present results must be
considered together with several limitations.

First, the calibration and evaluation datasets used in this study
consist only of Erd\H{o}s--R\'enyi graphs. This graph model was selected because
it provides a controlled setting in which the number of nodes and graph
density can be varied systematically. The resulting GBR models have therefore
been validated only for this graph distribution and should not be assumed to
produce the same results for other graph families without retraining and
evaluation on representative data.
This limitation concerns the models trained in the present study rather than
the penalty-tuning workflow itself. The implementation is modular and is
available in the project GitHub repository. The graph family is defined in the
\texttt{graph\_generation.py} module, which can be replaced with another graph
generator or dataset while leaving the remaining calibration, training, and
evaluation pipeline unchanged. The models can then be retrained for the selected
graph family. Evaluating the framework on additional synthetic graph families and
established real-world benchmark datasets is therefore an important direction for
future work.

A further limitation concerns the selected experimental baselines. The study
compares QA~HS mainly with Metis and, during the calibration stage, with
Kernighan--Lin. These methods were selected because they operate directly on
the graph-partitioning problem and do not depend on the QUBO penalty parameter.
The penalty intervals used in this study were obtained from the results of the
D-Wave hybrid solver and are therefore specific to this solver. Applying the same
penalty values to simulated annealing, tabu search, or another QUBO solver would
allow a comparison under the same formulation, but it might not show the best
performance of those methods because their preferred penalty values and solver
settings may differ. A fairer comparison would therefore require separate tuning
for each solver and comparable computational budgets. For Gurobi, an additional
comparison could be made with a constrained formulation in which the balance
condition is imposed directly rather than through a penalty term. We leave such a
broader evaluation for future work.

Scalability is also affected by the construction and storage of the QUBO
matrix. Although the cut contribution can be generated from a sparse
adjacency representation, the quadratic balance penalty introduces
interactions between every pair of nodes. The resulting BQM therefore
contains \(\mathcal{O}(n^2)\) quadratic coefficients. The experiments in
this study were limited to graphs with at most \(4000\) nodes, and
larger instances may require substantial memory and construction time.
Vectorized, block-based, and parallel coefficient generation may reduce
the construction time, but they do not eliminate the quadratic storage
requirement. Alternative formulations, including constrained quadratic
models or decomposition methods, should therefore be investigated for
larger graphs.

The experimental workflow additionally depends on D-Wave's proprietary
cloud platform and hybrid solver. The internal decomposition strategy and
use of quantum resources are not fully accessible to the user, and
execution time, solver availability, and financial cost depend on the
external service. Future work should therefore evaluate the proposed
penalty-tuning method with additional quantum, quantum-inspired, and
classical QUBO solvers.

The direct-QPU experiments were included as a complementary evaluation on
small graph instances, while the main penalty-tuning experiments focused on
QA~HS and larger graphs. The QPU runs used the default hardware settings, and
no sensitivity analysis of chain strength, annealing time, number of reads, or
embedding choice was performed. The reported results should therefore be viewed
as baseline results under the default configuration.

Further, machine learning could also be used to predict a suitable chain strength
for direct-QPU execution. A regression model could use features describing the
problem topology, QUBO coefficients, selected embedding, and target hardware.
Since the appropriate chain strength depends on both the embedding and the
hardware, such a model would require a separate dataset containing embedding
information, chain-break statistics, and solution-quality measurements.

Finally, the present study focuses on static bipartitioning. The general
framework could be extended to \(k\)-partitioning through a dedicated
multi-partition QUBO formulation or recursive application of
bipartitioning. However, such an extension would introduce additional
constraints and would require new calibration and validation experiments.
Dynamic environments, such as changing road networks or evolving
communication systems, would additionally require efficient model
updating and repeated optimization. These extensions represent promising
directions for applying data-driven penalty selection to more complex
engineering problems.

\section{Conclusion}
\label{sec:Conclusion}

This paper addressed the practical challenge of selecting the penalty
parameter in the QUBO formulation of the Minimum Bisection Problem, which
is commonly handled through heuristic estimates or repeated manual
testing.

We first derived an objective-based initial estimate of the penalty
parameter from the expected contribution of the cut objective. We then
refined this estimate by establishing graph-dependent lower and practical
upper bounds using properties obtained directly from the input graph.
We then introduced a machine learning approach based on two Gradient Boosting Regressors to
predict an effective interval of penalty multipliers from the number of
nodes, graph density, and the graph-dependent initial penalty estimate. The midpoint
of the predicted interval was used to calculate the final penalty
parameter and construct the corresponding QUBO model.

The proposed framework was evaluated on 126 independently generated
Erd\H{o}s--R\'enyi graphs with up to \(4000\) nodes. Under the adopted
experimental setup, the predicted penalty values enabled the D-Wave hybrid
solver to return balanced partitions for all evaluation instances and
lower cut values than PyMetis in every case. The results show that data-driven penalty selection can reduce the need for repeated parameter testing on new graph instances and make the MBP QUBO formulation more reliable. These findings apply to the graph distribution and solver configuration examined in this study.

Our future work will focus on evaluating the modular framework on
additional synthetic graph families and established real-world benchmark
networks. We also plan to compare alternative regression models and QUBO
solvers under matched experimental settings and to investigate more
scalable QUBO construction methods, constrained formulations, and
decomposition approaches for larger instances. Other promising directions include machine
learning-based chain-strength prediction for direct-QPU execution and
extensions to \(k\)-partitioning and dynamic networks.

\section*{Acknowledgment}

This work was supported by the Slovak Research and Development Agency
under project APVV-23-0512, by the project \textit{Research and Development of
Advanced Artificial Intelligence Solutions for Cyber Threat Detection
and Protection against Sophisticated Attacks}, Project Code
401101C360, and by the Scientific Grant Agency of the Ministry of
Education, Research, Development and Youth of the Slovak Republic and
the Slovak Academy of Sciences under project VEGA 1/0685/23.

During the preparation of this manuscript, the authors used AI-assisted tools solely to improve language, grammar, and readability,
including the wording of source-code comments. The tools were not used to
formulate the research methodology, design the experiments, implement
the algorithms, generate or analyze the experimental data, interpret the
results, or draw the conclusions. All scientific content was reviewed
and verified by the authors, who take full responsibility for the
published work.

Icons used in the graphical materials were obtained from
\url{https://www.flaticon.com}.

\section*{Code Availability}
\label{sec:code}

The source code, scripts used to generate the graph instances, trained
Gradient Boosting Regressor models, and experimental configurations are
publicly available in the following GitHub repository:
\url{https://github.com/rusnakrenata/qambp}.

The repository also contains instructions and software requirements
needed to reproduce the graph-generation, model-training, and evaluation
workflow. The original datasets used in this study are not distributed directly
but can be regenerated using the provided scripts and experimental
configurations. Access to the D-Wave Systems' quantum and hybrid solvers requires
an appropriate D-Wave Leap account.

\bibliographystyle{elsarticle-num}   % numeric style suitable for EAAI
\bibliography{bibliography}          % your .bib file name (without .bib)

\end{document}